\documentclass[twocolumn,prl,aps,showpacs]{revtex4}

\usepackage{graphicx}
\def\be{\begin{equation}}
\def\ee{\end{equation}}
\def\bea{\begin{eqnarray}}
\def\eea{\end{eqnarray}}
\def\ba{\begin{array}}
\def\ea{\end{array}}
\def\bdm{\begin{displaymath}}
\def\edm{\end{displaymath}}

\begin{document}

\title{Bose-Einstein Condensation in the presence of an artificial spin-orbit interaction}

\author{S.-K. Yip}

\affiliation{Institute of Physics, Academia Sinica, Nankang, Taipei
115, Taiwan}

\date{\today }

\begin{abstract}

Bose-Einstein condensation in the presence of a synthetic
spin-momentum interaction is considered, focusing on the case where
a Dirac or Rashba potential is generated via a tripod scheme. We
found that the ground states can be either plane wave states or
superpositions of them, each characterized by their unique density
distributions.

\end{abstract}



\maketitle

The possibility of producing an artificial gauge field \cite{Lin09}
opens up a new era in cold atom physics.  The concept of gauge
fields is ubiquitous in many branches of physics, from coupling of
electromagnetic fields to charged particles \cite{Sakurai} relevant
to ordinary laboratory settings to fundamental forces between
elementary particles \cite{Itzykson}.  In addition to the abelian
gauge field already realized in \cite{Lin09}, there are numerous
proposals to generate other artificial abelian and non-abelian gauge
fields, with and without optical lattices, with the prediction of
some exotic properties
\cite{Ruseckas05,Stanescu05,Stanescu08,Juzeliunas08,VC08,Goldman09,Larson09,Lim10,note}.


Two particular cases have captured much attention due to their
relation to condensed matter physics:  a Dirac-like term in the
Hamiltonian and Rashba interaction
\cite{Ruseckas05,Stanescu05,Stanescu08,Juzeliunas08,VC08,Goldman09,Larson09}.
The former is relevant to systems such as graphene \cite{Neto09},
whereas the latter to surface states in topological insulators
\cite{Hasan10} and is also discussed frequently in the context of
spintronics \cite{Nagaosa}, and both of them in non-centrosymmetric
superconductors \cite{NSC}.

While Dirac and Rashba interaction are often investigated in
fermionic electronic systems mentioned above, in this paper we
investigate the consequences of Bose-Einstein condensation (BEC) of
{\em Bosons} in such artificial gauge fields. We investigate the
possible ground states of this system, and show that one can have
BEC into plane-wave states or states corresponding to a
superposition of plane waves, depending on the physical parameters.
 We further investigate the observable
consequences, assuming that the gauge fields are produced via the
tripod scheme proposed in
\cite{Ruseckas05,Stanescu05,Juzeliunas08,VC08}. We shall see how the
single particle terms due to the laser fields or external potentials
and the interparticle interactions determine which ground state
would be realized.  Each ground state is associated with a
characteristic density distribution of the physical states involved
in the tripod scheme, enabling one to distinguish these states
experimentally.

  We mention here the closely related works of \cite{Stanescu08,Ho10,Wang10}.
\cite{Ho10} investigates BEC in the artificial gauge field in the
experiment cited in \cite{note}.  \cite{Wang10} considers also the
Dirac/Rashba interaction as in the present work, but without
reference to a particular production scheme and so did not discuss
the potentials arising from the laser fields, nor the density
distributions in the original atomic states.  They also limit
themselves to a less general form of the interparticle interaction
included in this paper. \cite{Stanescu08} did not include the
single-particle and interaction energies to be discussed in detail
below. They obtained a ground state qualitatively different from
\cite{Ho10,Wang10} and ours.

For definiteness, we consider the tripod scheme proposed in
\cite{Juzeliunas08}.  For convenience we shall first review it
briefly. We consider a cloud which is either three-dimensional or
confined by y-dependent optical potentials so that one only needs to
consider motion along x and z. Three (almost) degenerate states
$|1>$, $|2>$, $|3>$ are coupled to another state $|0>$ with light
fields with strengths $\Omega_{1,2,3}$.  That is, we have the
coupling term $H_{\Omega} = - |0> ( \Omega_1 <1| + \Omega_2 <2| +
\Omega_3 <3|)
  + c.c.$.  
  $\Omega_{1,2}$ are chosen to be
  plane waves of equal strengths propagating in opposite directions,
  while $\Omega_3$ involves a plane wave propagating perpendicular
  to the previous two with the same wavelength:  we have
  $\Omega_1 = \frac{1}{\sqrt{2}} \Omega  {\rm sin} \theta e^{- i k_0 x}$,
  $\Omega_2 = \frac{1}{\sqrt{2}} \Omega  {\rm sin} \theta e^{+ i k_0 x}$,
  and $\Omega_3 = \Omega {\rm cos} \theta e^{i k_0 z}$, with $\Omega$ and $\theta$
  parameters specifying the coupling strengths.  There are two
  linear combinations among the states $|1>-|3>$ which are not
  affected by $H_{\Omega}$.  They are referred to as "dark states"
  which we shall choose to be
  \bea
  |D_1> &\equiv& \frac{1}{\sqrt 2} \left[ |1> e^{+ i k_0 x} - |2> e^{- i k_0
  x} \right] e^{ i k_0 {\rm cos} \theta' z} \label{D1} \\
   |D_2> &\equiv& \frac{1}{\sqrt 2} {\rm cos} \theta \left[ |1> e^{+ i k_0 x} + |2> e^{- i k_0
  x} \right] e^{ i k_0 {\rm cos} \theta' z} \nonumber \\
   & & \qquad \qquad
    - {\rm sin} \theta |3>  e^{ - i k_0 ( 1 - {\rm cos} \theta') z} \label{D2}
    \eea
   Here we have generalized slightly \cite{Juzeliunas08} and include
  a yet undetermined parameter $\theta'$ in the z-dependent phase factors in eq
  (\ref{D1}) and (\ref{D2}).  The value for $\theta'$  does not
  affect the actual physics, but we shall choose a special value for it
   for convenience below.  The "bright"
  state $|B>$, orthogonal to both $|D_{1,2}>$, couples to $|0>$ with
  strength $\Omega$.  We shall assume that the energy splitting of these two
  resulting states from $|D_{1,2}>$ are large compared with all
  other energy scales considered below, hence
  $|D_{1,2}>$ are the only states  physically relevant.

   Within the $|D_{1,2}>$ subspace, the single particle Hamiltonian
   is given by

   \be
   {\bf H} = \frac{1}{2m} \left( \frac{\hbar}{i} \vec \nabla -
   \vec {\bf A} \right)^2 + {\bf \Phi} + {\bf V} \label{H1}
   \ee
    Here the matrix $\vec {\bf A}$ has elements $ \vec A_{\mu \nu}
   =  i < D_{\mu} | \vec \nabla D_{\nu} > $ ($\mu$, $\nu$ $=1$ or $2$)
   is the resulting
   (non-Abelian) gauge field, and $\Phi_{\mu \nu} =
   \frac{1}{2m} < \vec \nabla D_{\mu} | B > \cdot < B | \vec \nabla
   D _{\nu} > $ \cite{Ruseckas05}.  ${\bf V}$ includes the
   terms that may arise if the internal energies or the potentials acting on
the states $|1-3>$ are not identical.  We shall provide more details
 below. $\vec {\bf A}$ has components $\vec A_{11} = - k_0
{\rm cos} \theta' \hat z$, $\vec A_{12} = \vec A_{21} = - k_0 {\rm
cos} \theta \hat x$, $\vec A_{22} = - k_0 ( {\rm cos} \theta' - {\rm
sin}^2 \theta) \hat z$. With the choice ${\rm cos} \theta'= \frac{
{\rm sin}^2 \theta}{2}$, we then have

\be \vec {\bf A} = - k_0 {\rm cos} \theta \sigma_x \hat x
 - \frac{k_0}{2} {\rm sin}^2 \theta \sigma_z \hat z
 \label{A}
 \ee
where $\sigma_{x,y,z}$ are Pauli (pseudospin) matrices acting within
the $|D_1>$,$|D_2>$ space.
 With this $\vec {\bf A}$, we see that we have an anisotropic
 Dirac-like term in the Hamiltonian
 $ \frac{k_0}{m} \left( {\rm cos} \theta \sigma_x p_x +
  \frac{ {\rm sin}^2 \theta }{2} \sigma_z p_z \right)$,
 with the special case ${\rm cos} \theta
  = \sqrt{2} - 1 = \frac{ {\rm sin}^2 \theta } {2}$ where
  this term becomes isotropic in the x-z plane \cite{Juzeliunas08}.
  We shall however not assume such a special value for $\theta$ below.

  In the above vector potential $\vec {\bf A}$,  $\hat x$ and $\hat z$
  result from the $x$ and $z$ dependence of $\Omega_{1,2}$ and
  $\Omega_3$ respectively.  One can thus produce a Rashba-like term
  $ \frac{k_0}{m} \left( {\rm cos} \theta \sigma_x p_z -
  \frac{ {\rm sin}^2 \theta }{2} \sigma_z p_x \right)$ if one
  replaces $x$ by $z$ in $\Omega_{1,2}$ and $z$ by $-x$ in
  $\Omega_3$, that is, changing the directions of the lasers or
  relabeling the coordinates.  Alternatively, one can also use
  a different choice for $|D_{1,2}>$ corresponding to a
  "spin-rotation" \cite{VC08}.  For definiteness, we shall continue
  to deal with the Dirac-like term in our Hamiltonian.

  ${\bf \Phi}$ can be easily found to be
  $ \frac{k_0^2}{2 m}  {\rm sin}^2 \theta \left( \frac{ 1 + {\rm cos}^2 \theta}{2}
    + \frac{ {\rm sin}^2 \theta }{2} \sigma_z \right)$.  The first
    term is a momentum independent scalar which we shall drop.
    In the presence of a potential (or internal energies)
    $V_{1,2,3}$ on the states $|1-3>$, ${\bf V}$ is given by,
      apart from a constant which we shall again drop,
      ${\bf V} = \frac{1}{2} \left[ ( V_1 - V_2) {\rm cos} \theta \sigma_x
        + ( V_1 + V_2 - 2 V_3 )  \frac{ {\rm sin}^2 \theta }{2} \sigma_z
        \right]$. ${\bf \Phi}$ and ${\bf V}$ act like
         Zeeman fields
        in the space $|D_{1,2}>$.

        We shall first consider BEC in the presence of the gauge
        potential $\vec {\bf A}$, but ignoring ${\bf \Phi}$,
        ${\bf V}$, and interparticle interactions for the moment.
        These would be included later.
        The Hamiltonian is then simply the kinetic energy term
        ${\bf K} = \frac{1}{2m} (\vec p - \vec {\bf A})^2 $
        $= \frac{1}{2m} \vec p^2 - \frac{1}{m} \vec {\bf A} \cdot
        \vec p + \frac{1}{2m} \vec {\bf A}^2$.  $\vec {\bf A}^2$ is
        simply given by the constant $k_0^2 \left(
        {\rm cos}^2 \theta + \frac{ {\rm sin}^4 \theta }{4} \right)$, which
        we shall drop below.  ${\bf K}$ can easily be diagonalized
        giving the energies $E = \frac{p^2}{2m} \pm \frac{k_0}{m}
         \left[ {\rm cos}^2 \theta p_x^2 + \left(\frac{ {\rm sin}^2 \theta}{2} \right)^2
          p_z^2 \right]^{1/2}$.  BEC should occur in
        the state of the lowest energy, hence we make take the
        negative sign and find the $\vec p$ where
        $\frac{\partial E}{\partial p_x} = 0$ and
        $\frac{\partial E}{\partial p_z} = 0$.  Obviously $p_y$ is zero
        and for simplicity we shall not write this component explicitly.
         The possible minima are $\vec p = (0, \pm k_0 \frac{ {\rm sin}^2 \theta}{2} )
          \equiv ( 0, \pm p^0_z)$, with energy
         $E = - \frac{k_0^2}{2m} \left( \frac{ {\rm sin}^4
         \theta}{4} \right)$, and $\vec p = ( \pm k_0 {\rm cos}
         \theta, 0) \equiv ( \pm p_x^0, 0)$, with energy
          $E = - \frac{k_0^2}{2m} {\rm cos}^2 \theta$.  Hence if
          ${\rm cos} \theta < \sqrt{2} - 1$, the minima are at
          $\vec p = \pm p_z^0 \hat z$, whereas if
          ${\rm cos} \theta >  \sqrt{2} - 1$, the minima are at
          $\vec p = \pm p_x^0 \hat x$.  If ${\rm cos} \theta = \sqrt{2} -
          1$, we have the very special case that all momenta given
          by $(p_x^2 + p_z^2)^{1/2} = k_0 (\sqrt{2} -1 ) $ are
          degenerate.
          We shall not deal with this very special circumstance
          in the present paper.

            If we have condensation in one of these four minima,
            the corresponding wavefunctions in $|D_{1,2}>$ space are

            \bea
            \Psi_1(\vec r) &=& \Phi_1 e^{ i k_0 \frac{ {\rm sin}^2
            \theta}{2} z}
            \left( \ba{c} 0 \\ 1 \ea \right)  \label{P1} \\
            \Psi_2(\vec r) &=& \Phi_2 e^{ - i k_0 \frac{ {\rm sin}^2
            \theta}{2} z}
            \left( \ba{c} 1 \\ 0 \ea \right)   \label{P2} \\
           \Psi_3(\vec r) &=& \frac{\Phi_3}{\sqrt{2}} e^{ i k_0 {\rm cos} \theta x}
            \left( \ba{c} -1 \\ 1 \ea \right)   \label{P3} \\
            \Psi_4(\vec r) &=& \frac{\Phi_4}{\sqrt{2}} e^{-i k_0 {\rm cos} \theta x}
            \left( \ba{c} 1 \\ 1 \ea \right)   \label{P4}
            \eea
            where $\Phi_{1-4}$ are complex numbers.  We shall call them states 1-4
            (not be be confused with those which enter the tripod scheme).  We note here
            that even though these are plane wave states, they carry
            no current since we have demanded $\frac{\partial
            E}{\partial p_{x,z}} = 0$.  Alternatively, we should note that the
            velocity operator $\vec {\bf v}$ is given by
            $\frac{ \partial {\bf H}}{\partial \vec p}$, hence
            $\vec {\bf v} = \frac{1}{m} ( \vec p - \vec {\bf A})$
            $= \frac{\vec p}{m} + \frac{k_0}{m}
                \left( {\rm cos} \theta \sigma_x \hat x +
                 \frac{ {\rm sin}^2 \theta }{2} \sigma_z \hat z
                 \right)$.
                One can easily verify that the
                 expectation values of the velocity is zero in each
                 of the plane wave states (\ref{P1}-\ref{P4}) above.

    Since in each case we still have two degenerate minima, we must
    consider other terms in the Hamiltonian to determine which BEC
    would occur.  We first consider the other single particle terms in the
    Hamiltonian, that is ${\bf V} + {\bf \Phi} \equiv {\bf H}_h$
     $\equiv - \vec h \cdot \vec \sigma $, assuming for the
    moment that they are dominant over the interparticle
    interactions.  Due to the above discussion, we shall consider the case where ${\bf H}_h$
    has only $\sigma_{x,z}$ terms,  With finite $\vec h$, the relevant branch of the
    spectrum is
    \be
    E = \frac{p^2}{2m} - \frac{k_0}{m}
         \left[ \left({\rm cos} \theta p_x - \frac{m h_x}{k_0} \right)^2 +
         \left(\frac{ {\rm sin}^2 \theta}{2}  p_z - \frac{m h_z}{k_0}  \right)^2 \right]^{1/2}
         \label{Eh}
    \ee
    We now need to find the value of $\vec p$ where $E$ is a
    minimum.  In general this is very complicated for general $\vec
    h$.  In the limit of first order in $\vec h$, one can easily
    verified that the energies of the above four states become
    $E = - \frac{k_0^2}{2m} \left( \frac{ {\rm sin}^4
         \theta}{4} \right) \pm h_z$ and
         $E = - \frac{k_0^2}{2m} {\rm cos}^2 \theta \pm h_x$.
         Hence if $ {\rm cos} \theta < \sqrt{2} -1$ and $h_z < (>) 0$,
         then state 1 (2)
         should be realized,
    whereas if $ {\rm cos} \theta > \sqrt{2} -1$ and $h_x < (>) 0$,
    states 3 (4) should be realized.
    The above statements assumed that $\vec h$ is sufficiently small
    so that the ordering of energies are not changed except the lifting of
    degeneracies.   Also,
     strictly speaking, one also needs to
     consider the modification of the wavefunctions in eq
     (\ref{P1}-\ref{P4}) due to $\vec h$, as both the momentum and the "spin"
     wavefunction at which the minimum occurs are modified.  We
     however would not give these rather lengthy formulas here.
     (See however near the end of this paper below).

     Let us at this point give some physical properties related to these
     states,
     ignoring for simplicity
     the modifications of wavefunctions just mentioned.
     If $ {\rm cos} \theta < \sqrt{2} -1$ and $h_z <  0$,
         then (\ref{P1}) is the wavefunction in $|D_{1,2}>$ space.
     Using (\ref{D1}) and (\ref{D2}), one easily finds the
     wavefunctions in the original $|1-3>$ basis:

     \bea
     \psi_{1,2}(\vec r) &=& \frac{\Phi_1}{\sqrt{2}} {\rm cos} \theta
        e^{ \pm i k_0 x} e^{ i k_0  {\rm sin}^2 \theta z}  \\
     \psi_3(\vec r) &=& - \Phi_1 {\rm sin} \theta
        e^{ - i k_0 {\rm cos}^2 \theta z}
     \eea
    and so the corresponding particle densities are
    \bea
    |\psi_{1,2}(\vec r)|^2 &=& \frac{|\Phi_1|^2}{2} {\rm cos}^2
    \theta \label{d1a}\\
    |\psi_3(\vec r)|^2 &=&  |\Phi_1|^2 {\rm sin}^2 \theta
    \label{d1b}
    \eea
    Hence $|\psi_1(\vec r)|^2 = |\psi_2(\vec r)|^2$
    (again ignoring corrections
    due to $\vec h$, a statement which we shall not repeat).

    If $ {\rm cos} \theta < \sqrt{2} -1$ and $h_z >  0$, the system
    condenses into state 2 (eq (\ref{P2})) ,
    we have instead
    \bea
     \psi_{1,2}(\vec r) &=& \pm \frac{\Phi_2}{\sqrt{2}}
        e^{ \pm i k_0 x}   \\
     \psi_3(\vec r) &=&  0
     \eea
    and with the corresponding particle densities
    \bea
    |\psi_{1,2}(\vec r)|^2 &=& \frac{|\Phi_2|^2}{2} \label{d2a}\\
    |\psi_3(\vec r)|^2 &=&  0 \label{d2b}
    \eea
    which is very different from the state 1.

    For $ {\rm cos} \theta > \sqrt{2} -1$ and $h_x <  0$
    we have condensation into state 3, where
    \bea
     \psi_{1,2}(\vec r) &=& \frac{\Phi_3}{2}
     [ \mp 1 + {\rm cos} \theta]
        e^{ \pm i k_0  (1 \pm {\rm cos} \theta) x}
          e^{ i k_0 \frac{ {\rm sin}^2 \theta}{2} z}  \\
     \psi_3(\vec r) &=& - \frac{\Phi_3}{\sqrt{2}} {\rm sin} \theta
         e^{ i k_0   {\rm cos} \theta x}
        e^{ - i k_0 ( 1 - \frac{ {\rm sin}^2 \theta}{2}) z}
     \eea
     The densities are
    \bea
    |\psi_{1,2}(\vec r)|^2 &=& |\Phi_3|^2 \frac{( 1 \mp {\rm cos} \theta)^2}{4}
     \label{d3a} \\
    |\psi_3(\vec r)|^2 &=& |\Phi_3|^2 \frac{ {\rm sin}^2 \theta}{2}
    \label{d3b}
    \eea

For $ {\rm cos} \theta > \sqrt{2} -1$ and $h_x >  0$
    we have condensation into state 4.  We have
    \bea
     \psi_{1,2}(\vec r) &=& \frac{\Phi_4}{2}
     [ \pm 1 + {\rm cos} \theta]
        e^{ \pm i k_0  (1 \mp {\rm cos} \theta) x}
          e^{ i k_0 \frac{ {\rm sin}^2 \theta}{2} z}  \\
     \psi_3(\vec r) &=& - \frac{\Phi_4}{\sqrt{2}} {\rm sin} \theta
         e^{ - i k_0   {\rm cos} \theta x}
          e^{ - i k_0 ( 1 - \frac{ {\rm sin}^2 \theta}{2}) z}
     \eea
    with densities
    \bea
    |\psi_{1,2}(\vec r)|^2 &=& |\Phi_4|^2 \frac{( 1 \pm {\rm cos} \theta)^2}{4}
     \label{d4a} \\
    |\psi_3(\vec r)|^2 &=& |\Phi_4|^2 \frac{ {\rm sin}^2 \theta}{2}
    \label{d4b}
    \eea
    which are those of state 3 with $|1>$ and $|2>$ interchanged.
    In passing, we remark here also that, as indicated by the
    wavefunctions $\psi_{1,2,3} (\vec r)$ above, the atomic states
    $|1>$, $|2>$, $|3>$ are each associated with their
    characteristic wavevector, which can also be measured by
    time-of-flight experiments as in \cite{Lin09}.

    The above is for when the single particle terms dominate and
    determine the ground state.  Generally, if the energy
    differences between the above plane wave states are sufficiently
    small, we can have condensation into superposition of plane wave
    states.
     The energetics is analogous to that of superfluid
    mixtures \cite{mixtures} as we shall see.
    To be specific, let us consider the case where we have ${\rm cos}
    \theta < \sqrt{2} -1 $, with energies of states 1 and 2
    sufficiently lower than 3 and 4 so that the later two need not
    be considered.  We are left with considering condensation into
    states described by 1 and 2 in eq (\ref{D1}) and (\ref{D2}).
     (We are now considering normalized
    single particle states, that is,  without the $\Phi_{1,2}$
    coefficient).
    Consider the scattering between two particles, each one in the
    state 1.  Since the total momentum is $2 \vec p_0$,
    (here $\vec p_0 = ( 0, k_0 \frac{{\rm sin}^2 \theta}{2})$), the
    only relevant term in the low energy sector must correspond to
    that that the outgoing particles are also in state 1.
     Similarly, if there is one particle from each
    state 1 and 2, then the total momentum is zero and the relevant
    low energy process also correspond to outgoing particles with
    one each in state 1 and 2.  Let $d_{\vec p_0, 1}$ and $d_{-\vec p_0, 2}$ be
    the operators corresponding to the (normalized) plane wave states
    in states 1 and 2.  The general form of the Hamiltonian in the
    low energy subspace is
    $\frac{1}{2} \tilde{g}_{11} d_{\vec p_0, 1}^\dagger d_{\vec p_0,
    1}^\dagger d_{\vec p_0, 1} d_{\vec p_0, 1} +
     \tilde{g}_{12} d_{\vec p_0, 1}^\dagger d_{-\vec p_0,
    2}^\dagger d_{-\vec p_0, 2} d_{\vec p_0, 1} +
   \frac{1}{2} \tilde{g}_{22}  d_{-\vec p_0, 2}^\dagger d_{-\vec p_0,
    2}^\dagger d_{-\vec p_0, 2} d_{-\vec p_0, 2}$,
    where $\tilde{g}_{ij}$ are coefficients.    Other terms in the Hamiltonian are
    off-shell and can be dropped when we eventually take the
    mean-field approximation below.
    $\tilde{g}_{ij}$ can
    be evaluated once the interaction Hamiltonian in terms of the states
    $|1-3>$ in the tripod scheme is known.  There are two many possibilities
    so we do not provide the formulas here. We simply remark here
    that if the interaction among the particles $|1-3>$ are all
    identical, then $\tilde{g}_{11}=\tilde{g}_{12}=\tilde{g}_{22}$

    Within the mean-field approximation, we replace the operators
    $d_{\pm \vec p_0, 1 (2)}$ by the amplitudes $\Phi_{1,2}$, obtaining the
    interaction energy
    $E_{int} = \frac{1}{2} \tilde{g}_{11} |\Phi_1|^4
       + \tilde{g}_{12} |\Phi_1|^2 |\Phi_2|^2
         + \frac{1}{2} \tilde{g}_{22} |\Phi_2|^4$.
       Also, in the presence
         of the bias field ${\bf H_h}$, we have energy
         $E_h =  h_z (|\Phi_1|^2 - |\Phi_2|^2)$ to linear order in
         $\vec h$
    corresponding to the discussion below eq (\ref{Eh}).
    Note that there is no such term such as
    $ - h_x ( \Phi_1^* \Phi_2 + \Phi_2^* \Phi_1)$
     There cannot be a term in the form of
     $ - h_x ( d_{\vec p_0, 1}^\dagger d_{-\vec p_0, 2} +
           d_{\vec p_0, 2}^\dagger d_{-\vec p_0, 1})$ due to
      momentum conservation.

      Including also a chemical potential $\mu$ gives us
      then the total energy
      $ E_{tot} = - (\mu - h_z) |\Phi_1|^2 - (\mu + h_z) |\Phi_2|^2
        + E_{int}$ as in the case of a mixture of two species 1 and 2, with
        effective chemical potential for component 1 (2) being
        $ \mu_{1,2} = \mu \mp h_z$.  Note that therefore there are no
        terms which would depend on the phase difference between
        $\Phi_{1,2}$.  We shall have another point of view of this
        below.  The phase diagram of a two component mixture is
        well-known \cite{mixtures}, and we shall not repeat those
        results here.
        The state corresponding to a "mixture" between 1 and 2 has
        a wavefunction in the $|D_{1,2}>$ space as a
        the superposition of (\ref{P1}) and (\ref{P2}), that is,
          \bea
        \Psi(\vec r) &=& \Phi_1 e^{ i k_0 \frac{ {\rm sin}^2
            \theta}{2} z}
            \left( \ba{c} 0 \\ 1 \ea \right) +
             \Phi_2 e^{ - i k_0 \frac{ {\rm sin}^2
            \theta}{2} z}
            \left( \ba{c} 1 \\ 0 \ea \right)  \label{Pm12}
            \eea
       and
        likewise for their projection into the states $|1-3>$.
       We can understand easily why the energy is independent of
       the relative phase between $\Phi_{1,2}$.  Changing the
       relative phase between these two complex numbers can simply
       be reabsorbed by a shift of the origin for $z$.  Below, we
       shall for simplicity assume that both $\Phi_{1,2}$ are real.

       The densities for this "mixture" of 1 and 2 are
         \begin{widetext}
        \bea
        |\psi_{1,2}(\vec r)|^2 &=&
        \frac{1}{2} \left[  {\rm cos}^2 \theta |\Phi_1|^2 + |\Phi_2|^2
           \pm 2 \Phi_1 \Phi_2 {\rm cos} \theta {\rm cos}
           ( k_0 {\rm sin}^2 \theta z ) \right] \label{dm12a} \\
        |\psi_3 (\vec r)|^2 &=& |\Phi_1|^2  {\rm sin}^2 \theta
        \label{dm12b}
        \eea

        The coefficients of $|\Phi_{1,2}|^2$ are the same as eq
        (\ref{d1a})-(\ref{d2b}).
        An interference term $\propto \Phi_1 \Phi_2 $ arises
        from the wavevector difference $k_0 {\rm sin}^2 \theta
        \hat z$ between the $\Phi_{1,2}$ components in eq
        (\ref{Pm12}).
        Similar discussion holds for a "mixture" of states 3 and 4.
        The densities are

         \bea
        |\psi_{1,2}(\vec r)|^2 &=&
        \frac{1}{4} \left[ (1 - {\rm cos} \theta)^2 |\Phi_3|^2 + (1 + {\rm cos} \theta)^2 |\Phi_4|^2
          - 2 \Phi_3 \Phi_4 {\rm sin}^2 \theta {\rm cos}
           ( 2 k_0 {\rm cos} \theta x ) \right] \label{dm34a} \\
        |\psi_3 (\vec r)|^2 &=& \frac{1}{2} {\rm sin}^2 \theta  \left[ |\Phi_3|^2 + |\Phi_4|^2
         + 2 \Phi_3 \Phi_4  {\rm cos} ( 2 k_0 {\rm cos} \theta x )
         \label{dm34b}
        \right]
        \eea
        \end{widetext}
        The densities thus in general oscillate in space, as has
        been discussed in \cite{Ho10,Wang10}.

        Next we would like to consider in more detail the case of ${\bf H}_h
        = - h_z \sigma_z$, without restricting to small $h_z$.  Recall that
        $\hat {\bf \Phi}$ contains such a component even in the special case
        that ${\bf V} = 0$, and also that $h_z < 0$ in that case.
       Due to the lack of space we would
        mainly just state the results.  First we ignore interparticle
        interactions. If ${\rm cos} \theta <
        \sqrt{2}-1$, the state 1 (2) with $\vec p = ( 0, \pm k_0
        \frac{ {\rm sin}^2 \theta}{2} )$ is the energy minima  if $h_z <
        (>) 0$ with energy
        $E = - \frac{k_0^2}{2m} \left( \frac{ {\rm sin}^4
         \theta}{4} \right) -  |h_z|$.
         For ${\rm cos} \theta > \sqrt{2}-1$, and if $|h_z| < h_c$ where
         $h_c \equiv \frac{k_0^2}{m} D(\theta)$
         with $D(\theta) \equiv {\rm cos}^2 \theta - \left(
         \frac{{\rm sin}^4 \theta}{4} \right)$ (note $D>0$ in
         the above range for $\theta$), then the energy
         minima occurs at $(\pm p_x^0, p_z^0)$ where
         $p_x^0 = k_0 {\rm cos} \theta [ 1 -
         (\frac{\tilde h_z}{D})^2 ]^{1/2}$ and
         $p_z^0 = - \tilde{h}_z \frac{{\rm sin}^2 \theta}
         {2 D}$.  Here $ \tilde h_z \equiv h_z/(k_0^2/m)$ is a dimensionless
         measure of the energy $h_z$.
         Note there are two degenerate minima of opposite $p_x$, and
          $sgn(p_z^0) = - sgn(h_z)$. We shall call these states 5 and 6.
          The energy is given by
          $E = - \frac{k_0^2}{2m}  \left[ {\rm cos}^2
         \theta + \frac{\tilde{h}_z^2}{D} \right] $.
         At $\mp h_c$, $\vec p$ becomes $( 0, \pm k_0 \frac{{\rm
        sin}^2 \theta}{2} )$ and thus merge with the other minima
        stated before.  For $|h_z|> h_c$, the minima is at
        $ \vec p = ( 0, \pm k_0 \frac{{\rm
        sin}^2 \theta}{2} )$ according to $h_z < (>) 0$.

       It is convenient to define the quantity $\beta
       \equiv {\rm tan}^{-1} \left[ \frac{-\tilde h_z /D} { 1 - (\tilde h_z
       /D)^2} \right]$.  $\beta = 0$ if $h_z = 0$ and $\beta =
        - sgn(h_z) \frac{\pi}{2}$ at $|h_z| =   h_c$.
        The wavefunction in $|D_{1,2}>$ space for state 5, with wavevector $(p_x^0, p_z^0)$,
        is
        \be
        \Psi_5 (\vec r) = \frac{\Phi_5}{\sqrt{2}} e^{ i (p_x^0 x +
        p_z^0 z)}  \left(
        \ba{c} - ({\rm cos} \frac{\beta}{2} - {\rm sin} \frac{\beta}{2}) \\
                   {\rm cos} \frac{\beta}{2} + {\rm sin} \frac{\beta}{2} \ea
                   \right)
        \label{P5}
        \ee
        whereas that for state 6 with  wavevector $(-p_x^0, p_z^0)$
        is
           \be
        \Psi_6 (\vec r) = \frac{\Phi_6}{\sqrt{2}} e^{ i (-p_x^0 x +
        p_z^0 z)}  \left(
        \ba{c}  {\rm cos} \frac{\beta}{2} - {\rm sin} \frac{\beta}{2} \\
                   {\rm cos} \frac{\beta}{2} + {\rm sin} \frac{\beta}{2} \ea
                   \right)
        \label{P6}
        \ee
          These two states are orthogonal due to their different
          wavevectors $(\pm p_x^0, p_z^0)$, but their "spinor"
          part has finite (${\rm sin} \beta$) overlap.
          At $h_z = 0$, $\beta=0$, $p_x^0 = k_0 {\rm cos} \theta$,
          $p_z^0 = 0$,  states 5 and 6 are identical
          respectively with state 3  and 4 .
          At $h_z = - h_c$, $\beta = \frac{\pi}{2}$, and as stated,
          $ p_x^0 =0$, $p_z^0 = k_0 \frac{{\rm
           sin}^2 \theta}{2}$, they become identical with each other
          and with state 1.

          The densities corresponding to state 5 are
          \begin{widetext}
          \bea
          |\psi_{1,2}(\vec r)|^2 &=& \frac{|\Phi_5|^2}{4}
          \left[ 1 + {\rm cos}^2 \theta
           - {\rm sin} \beta {\rm sin}^2 \theta
           \mp  2 {\rm cos} \theta {\rm cos} \beta \right]  \label{d5a} \\
          |\psi_{3}(\vec r)|^2 &=& \frac{|\Phi_5|^2}{2}
             {\rm sin}^2 \theta ( 1 + {\rm sin} \beta ) \label{d5b}
          \eea
          \end{widetext}
          The densities of state 6 are those of 5 with
          states $|1>$ and $|2>$ interchanged.

          For a superposition between states 5 and 6, the densities have three
          contributions, the terms proportional to $|\Phi_{5,6}|^2$
          are the same as those in eq (\ref{d5a}) and (\ref{d5b}).
          In addition, there are the interference terms
          \bdm
          -  \frac{\Phi_5 \Phi_6}{2} \left[ {\rm sin}^2 \theta -
          {\rm sin} \beta (  1 + {\rm cos}^2 \theta) \right]
          {\rm cos} (2 p_x^0 x)
          \edm
          for $|\psi_{1,2} (\vec r)|^2$, and
            \bdm
            \Phi_5 \Phi_6 {\rm sin}^2 \theta ( 1 + {\rm sin} \beta )
            {\rm cos} (2 p_x^0 x)
            \edm
            for $|\psi_{3} (\vec r)|^2$.
            Note that they reduce to the appropriate limits
             eq (\ref{dm34a}) (\ref{dm34b}) if $\beta = 0$.
             At $\beta = \frac{\pi}{2}$, we get eq (\ref{d1a}) and
             (\ref{d1b}) with $\Phi_1 \to \Phi_5 + \Phi_6$.

             If ${\bf V} = 0$, we have $h_z = - \frac{k_0^2}{m}
                 \left( \frac{{\rm sin}^2 \theta}{2} \right)^2$.
                 Hence $|\tilde h_z|  =  h_c$ at
                 ${\rm cos} \theta = \frac{ \sqrt{6} - \sqrt{2}}{2}
                   \approx 0.518$, corresponding to $\theta \approx
                    0.326 \pi$.  Hence, due to this finite
                    $h_z$, we have states 5, 6 or their mixtures if
                    $\theta < 0.326 \pi$, and state 1 if otherwise
          (assuming that interaction does not change the ordering of
           the energies).

   In \cite{VC08}, standing waves for $\Omega_{1,2}$ are considered
    instead of plane waves.  This situation can be treated in a similar manner
    as in this paper.  The corresponding results can be obtained by
    a simple unitary transformation among the atomic states $|1>$ and $|2>$.

            In conclusion, we have considered Bose-Einstein
            condensation in an artificial spin-orbit field.
            In general BEC occurs in a state with finite wavevector,
            or their superpositions.  Due to the momentum-pseudospin coupling,
            each state
            is characterized by their unique density distributions.

 This research was supported by the  National Science Council of
 Taiwan.
 The author would also like to thank
  the Aspen Center for
 Physics where this study was motivated.

\end{document}